\documentclass[10pt,a4paper]{article}  % Comment this line out if you need a4paper
\usepackage{cite}
\usepackage{amsmath,amssymb,amsfonts}
\usepackage{algorithmic}
\usepackage{graphicx}
\usepackage{textcomp}
\usepackage{geometry}
\usepackage{caption}
\usepackage{subcaption}
\usepackage{rotating}
\usepackage{adjustbox}
\usepackage{changepage}
\usepackage{lipsum} 
\usepackage{afterpage}
\usepackage{etoolbox}
\usepackage{acronym}
\appto\appendix{\addtocontents{toc}{\protect\setcounter{tocdepth}{0}}}
\usepackage{nomencl}

\newtheorem{assumption}{Assumption}

\appto\listoffigures{\addtocontents{lof}{\protect\setcounter{tocdepth}{1}}}
\appto\listoftables{\addtocontents{lot}{\protect\setcounter{tocdepth}{1}}}
\title{\LARGE \bf Accurate EEG-based linear state-space model of cortical
dynamics during a passive motor task with exogenous input\thanks{The authors are with departments of BioMechanical Engineering and Delft Center for Systems and Control, Faculty Mechanical Engineering, Delft University of Technology, The Netherlands. Email: {\tt\small sanna.bakels@gmail.com}, {\tt\small m.l.vanderuit-1@tudelft.nl}, {\tt\small m.jafarian@tudelft.nl}. This research has been conducted under the co-lead of M. van de Ruit and M. Jafarian.}}
\author{Sanna Bakels, Mark van de Ruit, and Matin Jafarian}

\date{}

\begin{document}
\maketitle

\begin{abstract}
Objective. Revealing the dynamics of the brain in the presence of exogenous inputs is crucial for mechanistic understanding, interventions and regulations. We study state-space mathematical modeling of brain's cortical dynamics using electroencephalography (EEG) data in an experiment with continuous exogenous input. Approach. We use subspace identification techniques to model the brain dynamics in a passive motor task experiment. The EEG data were recorded while participants were seated with their wrist strapped to a haptic manipulator. The manipulator imposed a continuous multisine angular perturbation to the wrist as the exogenous input to the brain. Main result. Using two subspace identification algorithms, we show that linear time-invariant state-space models accurately represent the measurements, in a latent space, assuming that the EEG data are the models' output. The model is verified and validated using data from seven participants. Moreover, we construct linear maps to relate the latent space dynamics to the neural source space. We verify that the findings of our model align with those identified in previous studies. Significance. Accurate linear model in the presence of exogenous input provides a pathway for designing parsimonious and simple controllers. In addition, it enables development of online monitoring, connectivity analysis and control relevant to neurorehabilitation, motor learning, and brain-computer interfacing.  
\end{abstract}
%%%%%%%%%%%%%%%%%%%%%%%%%%%%%%%%%%%%%%%%%%%%%%%%%%%%%%%%%%%%%%%%%%%%%%%%%%%%%%%%%%%%%%%%%%%%%%%%%%%%%%%%%%%%%%%%
\section{Introduction}
\label{sec:introduction}
The human brain is an incredibly complex organ characterized by billions of neurons and trillions of synaptic connections, which serve as junctions for neuronal intercommunication. This intricate neural network is the foundation for the brain's ability to process information, enabling cognitive functioning and motor activities \cite{vanKordelaar2021TheStroke}. Understanding the dynamics of large-scale brain networks is crucial for advancing both theoretical neuroscience and practical applications such as brain interface devices and clinical diagnostics.

Brain imaging techniques such as functional magnetic resonance imaging (fMRI) and electroencephalography (EEG) visualize the collective activity of thousands of neurons active at the same time. Studying the simultaneous or subsequent activity of brain areas or networks enables the study of information processing across the brain. Although fMRI is better at providing the location of brain activity, the high temporal resolution makes EEG an attractive option for researchers seeking to model the temporal dynamics of brain networks \cite{Michel2012TowardsTool}. Moreover, EEG stands out as a versatile and cost-effective solution, advantageous for use in environments where participants may not be restricted to stationary positions \cite{Do2021HumanNavigation}. The development of EEG-based dynamic models of large-scale brain activity \cite{vakitbilir2025multivariate} facilitates an in-depth understanding of cerebral functioning and intricate communication dynamics between different neural sources \cite{Breakspear2017DynamicActivity}. 

The conventional mathematical framework to model brain dynamics using {EEG} data relies on the utilization of linear vector autoregressive, namely {VAR} functions \cite{Cheung2010EstimationModels,Parra2005RecipesEEG}. These models have demonstrated their ability to classify brain activity \cite{Anderson1998MultivariateTasks, Chisci2010Real-TimeMachines}. However, time-series data originating from applications subjected to filtering, down-sampling, and noise, as with neurophysiological data, often benefit from a moving-average (MA) component. The MA component captures how the present value of the time series is influenced by past errors over a specific window in time. This is not efficiently captured by a finite-order VAR model \cite{Barnett2015GrangerModels}.  To address this issue, the vector autoregressive moving average model (VARMA) has been used instead of {VAR} models, to capture the idea that past disturbances in the system can still have an effect on the current state of the system \cite{Barnett2015GrangerModels, Ubeyli2010LeastSignals}. A drawback of VARMA is its increased complexity compared to simpler models, such as VAR, as it is prone to challenges in identification, estimation, and specification. These difficulties arise mainly due to the flexibility of the model structure, leading easily to over-parametrization if care is not exercised. 

While the above-mentioned techniques are useful in studying the brain's connectivity, more advanced goals such as prediction and control call for more advanced frameworks. State-space models present a promising alternative, offering a more comprehensive mathematical framework to capture both the autoregressive and moving-average characteristics of EEG data. In addition, these models can incorporate external stimuli, providing both a better understanding of brain connectivity during tasks and the possibility of prediction and control. Furthermore, the capacity of state-space models to accommodate missing data is particularly advantageous in neuroscience, where data quality and availability can be variable and inconsistent.

State-space models have been developed to capture the brain's dynamics from {EEG} \cite{Cheung2010EstimationModels,Manomaisaowapak2022GrangerApproach} or {fMRI} data \cite{Becker2018Large-ScaleIdentification,dziarnowska2026associative}, given the established equivalence between the {VARMA} model and state-space model \cite{Barnett2015GrangerModels}. 
To date, research has mainly focused on applying state-space models to assess brain connectivity in input-free EEG datasets, particularly in resting-state experiments where no explicit external stimuli are present. This paper addresses a critical gap by incorporating exogenous inputs into the state-space framework, enabling the modeling of brain networks under externally driven conditions. 
 
Subspace identification algorithms~\cite{Verhaegen2007FilteringApproach} emerge as a promising data-driven technique to construct linear time-invariant  state-space models out of EEG data. These methods unveil the hidden characteristics of the underlying system directly from the input-output measurements without requiring an exhaustive understanding of the system's physical equations. Unlike several other identification algorithms, such as the prediction error method, subspace identification eliminates the need for model parameterization \cite{Verhaegen2007FilteringApproach}. Additionally, the system model is obtained non-iteratively, improving computational efficiency and reducing model complexity \cite{Verhaegen2007FilteringApproach}. Moreover, applying subspace identification for state-space model identification opens the path to extracting the brain's input-dependent connectivity from the model. 

In this work, first, we employ two subspace identification algorithms to obtain a linear state-space model for EEG data recorded while participants receive continuous sensory input. This is achieved by developing a framework including EEG pre-processing and constructing linear models, subject to additive ergodic white noise in output and dynamics, using two well-celebrated subspace identification algorithms, namely PO-MOESP \cite{Verhaegen2007FilteringApproach} and N4SID \cite{VanOverschee1994N4SID:Systems}. We employ these techniques to construct linear models in a latent space, assuming that the EEG data are the models' output. We then discuss the validity of the models obtained and compare their performance. Our results show that the PO-MOESP algorithm leads to a linear and time-invariant model with low estimation error, that is, the VAF (Variance Accounted For) value reaches $86.7\%$ and the RMSE (Root Mean Squared Error) is around $0.2$. Our results suggest that a linear time-invariant model is able to represent the dynamics of the brain network while participants receive a continuous external sensory disturbance to the wrist. 

Moreover, we construct linear maps, using a Lead-Field and a transformation matrix, to establish a relation between the obtained model in the latent space and activities of neural sources. We verify that the source activity found by our model is primary located in the contralateral sensorimotor areas aligning with findings in a previous fMRI study \cite{szameitat2012cortical}. We then discuss the implications of our findings and the prospects of using the proposed approach to design controllers.

This paper is organized as follows. The methods, including problem formulation, the proposed algorithm, data processing, sub-space identification, and latent to source mapping are presented in Section~\ref{sec:pf}. Results of sub-space-based modeling in the latent space as well as verification by mapping from the latent space to source space are presented in Section~\ref{sec:re}. Finally, Section~\ref{sec:dis} presents discussion about our findings, their implications, and conclusions.

%%%%%%%%%%%%%%%%%%%%%%%%%%%%%%%%%%%%%%%%%%%%%%%%%%%%%%%%%%%%%%%%%%%%%%%%%%%%%%%%%%%%%%%%%%%%%%%%%%%%%%%%%%%%%%%%%%%%%%%%%%%%%%%%%%%%%%%%%%%%%%%%%%%%%%%%%%%%%%%%%%%%%%%%%%%%%
\section{Methods}\label{sec:pf}
In this section, we first describe the experiment during which EEG data were collected. Then, we continue to formulate the problem, research questions, and propose our approach to solving the problem. Moreover, we provide an overview on sub-space identification algorithms, and the mapping from the latent space to source space.
%================================================================================================
The problem addressed in this study is how to obtain a state-space model for brain dynamics using EEG recordings in a task in which participants receive an identifiable exogenous input. 
Effectively, this is, given the collected EEG data, how to obtain a state-space model in the general form
$$z(k+1)= A_z z(k) + B_z u(k) + w (k)$$ $$y(k)=C_z z(k)+ D_z u(k)+ v(k)$$
\noindent where $z$ represents the source activity corresponding to neural activity, and $y$ corresponds to the output representing the electrical activity recorded by the electrodes placed on the scalp during {EEG} measurements. The matrices $A_z, B_z, D_z$ represent the system dynamics, while $w, v$ are noise terms associated with the source signal and the output measurements, respectively. Subspace identification allows us to construct the model matrices assuming that the state is observable from the output. However, the `state' in this modeling approach is not necessarily equal to the state variable of the cortical regions, which contribute to the generation of the outputs. In fact, we first use the measurements to construct a linear time-invariant model in a latent space. Then, a linear operator is used to map the latent space to the source space. In Section~\ref{sc3:subspaceID}, we present the modeling and our approach to solve the above problem. The data requirements, choice of experimental setup, and steps of EEG signal processing are discussed below.

\subsection{Data and input requirements}\label{sc3:experimentalFramework}
Sustained steady-state responses (SSR), evoked by continuous inputs, play a vital role in ensuring model stability. A task that exhibits a stable and synchronized rhythmic pattern is crucial to maintain this stability during dynamic brain modeling \cite{Vlaar2017QuantificationStroke}.

To accurately model brain dynamics, it is necessary to effectively capture the neural responses during task engagement. This requirement aligns with a core principle of open-loop subspace identification methods, which rely on persistently exciting input to capture the full range of system dynamics. The notion of persistency of excitation requires that the input signal contains sufficient frequency content to excite all relevant system modes. This allows a comprehensive exploration of the behavior of the system for accurate parameter estimation \cite{Verhaegen2007FilteringApproach}. The selected data set shall meet these criteria by providing continuous, measurable input that elicits sustained brain oscillations, i.e., the consistent elicitation of measurable neural responses over time due to the continuous application of the periodic multi-sine input \cite{pintelon2012system}. To ensure this, the excitation signal was applied continuously over multiple periods, allowing the system to reach a steady oscillatory regime and enabling reliable estimation of the underlying dynamics.
%=====================================================================================
\subsubsection{Experimental setup\\[1mm]}
\hspace{-8mm} All data used in this study were obtained, with permission, from an experiment performed by Vlaar et al. \cite{Vlaar2017QuantificationStroke}. A concise overview of the experimental procedure and data collection is provided below; for a detailed description, the reader is referred to \cite{Vlaar2017QuantificationStroke}. The dataset comprises EEG recordings from healthy participants undergoing continuous robotic wrist manipulation, providing a quantifiable input while inducing steady-state brain responses. The resulting data satisfies the data requirements of the modeling approach used in this study. A video, corresponding to the experiment, is provided in the Supplementary Materials.

Ten unimpaired participants, aged 22 to 25, participated in the study. The study was approved by the local ethics committee. A robotic manipulator, the `Wristalyzer', MOOG Inc., Nieuw-Vennep, Netherlands, applied continuous periodic angular perturbations to the dominant wrist of each participant while they remained seated with their forearm secured. EEG was recorded at 2048 Hz using a TMSi Refa amplifier (TMSi, Oldenzaal, Netherlands) and an electrode cap with 64 Ag/AgCl electrodes, arranged according to a 10/10 system. A separate electrode (Blue Sensor N, Ambu, Ballerup, Denmark) was connected to the left mastoid process and served as the participant's ground. 

\begin{figure}[tb]
\centering
\includegraphics[width = 0.4\linewidth]{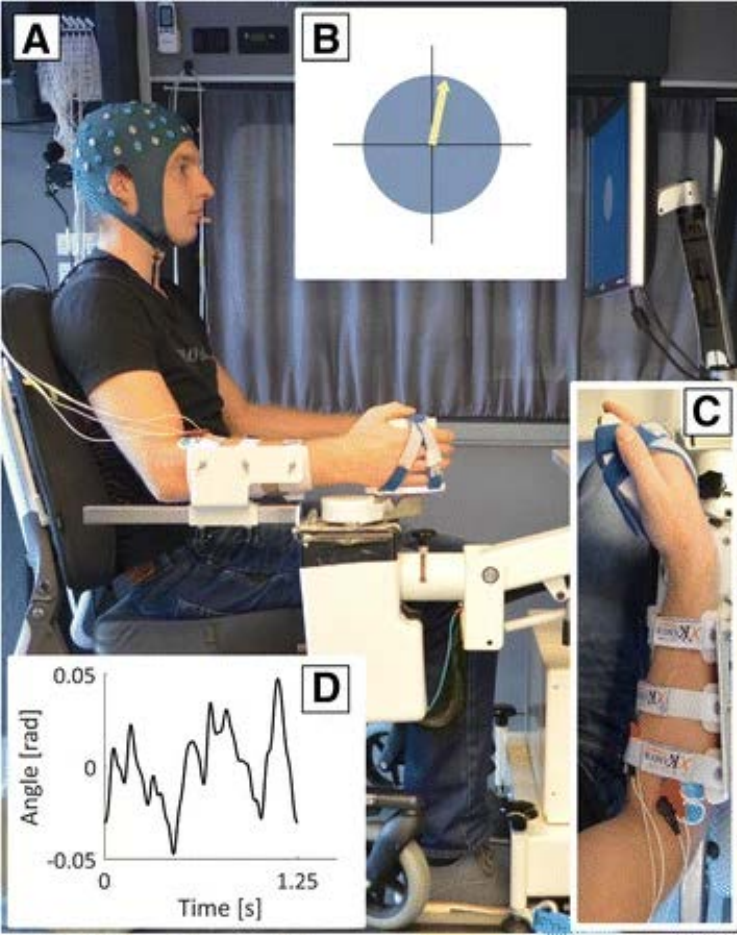}
\caption{Experimental setup from \cite{Vlaar2017QuantificationStroke} (Figure obtained with permission): (A) The forearm of the participant is strapped into an armrest, and the hand is strapped to the handle of the robotic wrist manipulator. (B) %The yellow arrow, only visible when the subject applies torque, is not relevant in the context of the passive execution of the task. 
The screen was not essential for this experiment and served only as a visual fixation point for the participants during the task. (C) Close-up of the arm in the robotic manipulator where the arm is placed in the neutral angle of $20\deg$ wrist flexion. (D) The plot of one complete cycle of the disturbance signal applied to the wrist.}
\label{fig3:experimentalSetup}
\end{figure}

Participants were instructed to relax their wrist and ignore the applied perturbations, a passive task. For this task, 20 trials of 12.5 s were recorded. The exogenous input signal consisted of periodic angular perturbations applied to the wrist joint by the robotic manipulator. The perturbation signal was designed \cite{Vlaar2017QuantificationStroke} as a multisine signal with random-phase characteristics, combining multiple sinusoids relevant to studying the wrist's dynamics. The random phase multisine input signal was defined as \cite{Vlaar2017QuantificationStroke}
$$
u(t) = \sum_{k=1}^{N} A_k \cos\left(2\pi f_k t + \phi_k\right),
$$
where $f_k = k f_0$ with $f_0 = 1/T$, $A_k \neq 0$ for the excited frequencies, and $\phi_k $  is the random phase at frequency $f_k$ which is taken from a uniform distribution. The signal excites frequencies in the range $0.8–19.2 Hz$ with a frequency resolution of $0.8 Hz$, balancing frequency resolution and the number of periods that can be recorded in a given measurement time. The angular disturbances were identical in magnitude and phase for all participants, and applied around a neutral wrist angle of flexion $20\deg$, with a root mean square deviation of $0.02$ radians.

This multisine design ensures persistency of excitation of sufficiently high order, as it provides broadband spectral content over the frequency range of interest. Compared to stochastic inputs such as white noise, multisine signals concentrate signal power at selected frequencies, improving SNR, enable noise reduction through period averaging.
%===================================================================================
\subsection{Proposed approach: overview} \label{sc3:algorithm} 
\begin{figure}[h!]
    \centering
    \includegraphics[width = 0.7\linewidth]{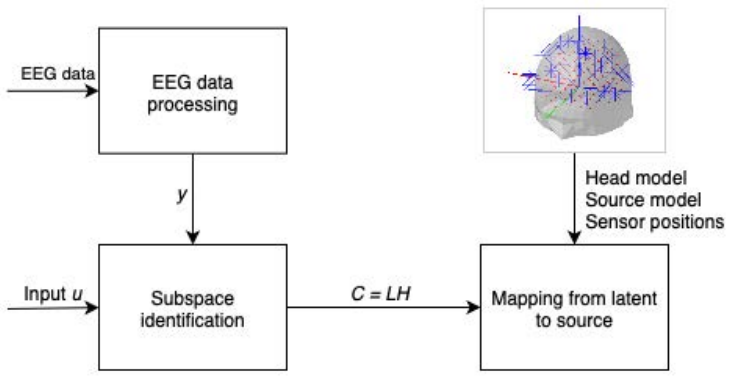}
    \caption[Block diagram of the designed framework for modeling brain dynamics]{Block diagram illustrating the algorithmic framework for building a dynamic system in an individual participant. Each block represents a distinct stage within the framework. The last block is detailed in Figure \ref{fig3:step3}.}
    \label{fig3:thesisMethod}
\end{figure}

Here, we provide an overview capturing all stages from data processing to modeling using a block diagram illustrated in Figure \ref{fig3:thesisMethod}. First, a state-space model is constructed based on the pre-processed and restructured EEG data. Second, two subspace identification algorithms, namely PO-MOESP and N4SID, will identify a linear state-space model subject to additive white noise. The last step entails converting state information obtained through subspace identification into discernible source signals by solving the EEG's inverse problem. This transformation facilitates the derivation of insights into interconnections among brain sources and regions from the recorded {EEG} data. 
All computations were performed using MATLAB R2023A (The MathWorks, Inc., Natick, MA, USA) alongside specialized toolboxes, including EEGLAB for EEG data processing \cite{Delorme2004EEGLAB:Analysis}, the System Identification Toolbox, the LTI System Identification Toolbox 2.4 \cite{Verhaegen2007FilteringSoftware}, and FieldTrip for lead-field matrix computations \cite{Oostenveld2011FieldTrip:Data}.
%------------------------------------------------------------------------------
\subsubsection{Processing EEG data} \label{sc3:preprocess}
Before proceeding to modeling, we explain the steps taken for EEG data processing. First raw {EEG} data is pre-processed and averaged to generate the {SSR}. These {SSR}s are the output $y$, serving as the fundamental basis for constructing a state-space model to represent the brain dynamics. 

The pre-processing of the dataset of \cite{Vlaar2017QuantificationStroke} was performed using standard filtering techniques to enhance the signal quality and isolate neural activity. The dataset used for analysis consisted of EEG recordings from seven participants, as three participants were excluded due to unsynchronized input signals or preprocessing discrepancies. 

First, the {EEG} data was filtered between $0.8$ and $120 Hz$ using a $4th$ order Butterworth filter. In addition, we used a band-stop filter, an infinite impulse response notch filter, around 50 Hz and 100 Hz to remove line noise and its harmonics. Any channels that did not present any recording activity (flat lines) were removed before the remaining signals were re-referenced to the common average.

Independent component analysis (ICA), a blind source separation technique, was applied using the Infomax algorithm \cite{Bell1995AnDeconvolution.}  as implemented in RUNICA in EEGLAB to isolate neural signals from artifacts, including muscle activity and eye blinks. The IClabel EEGLAB \cite{pion} plugin was used to remove components identified as muscle ($>80\%$ certainty), eye, heart, line noise, or channel noise (all $>90\%$ certainty). Any component identified as originating from a source other than the brain was removed, and the remaining components were linearly combined. The data was then segmented into 1.25-second epochs aligning to the periodic input signal, excluding the first and last epochs of each of the 20 trials to minimize transient effects. This process resulted in 160 epochs per participant.

The wrist angle data were also analyzed to exclude epochs in which the perturbation was not executed as designed. For each epoch, a z-score for the recorded wrist angle with respect to the designed wrist angle was calculated. An epoch was flagged as erroneous whenever the z-score exceeded 3. All erroneous epochs were subsequently excluded from both the wrist angle as well as the pre-processed EEG data. Finally, before proceeding to the state-space modelling, the {SSR} was calculated by averaging the EEG across the remaining epochs, thereby improving the signal-to-noise ratio. For each participant, this yielded a $1.25$-second signal per electrode, producing a dataset with dimensions $62 \times 2560$ for all seven participants.
%----------------------------------------------------------------------------------------------------------
\subsection{State-space modeling of brain's network dynamics in a latent space} \label{sc3:subspaceID}
This section presents building a sub-space-based linear model representing the inherent dynamics of the brain during the execution of the experiment explained in Section~\ref{sc3:experimentalFramework}. We explain the modeling process, which allows for estimating the brain's dynamics from the measured input and output data. In this context, the known components include the exogenous input $u$, related to the imposed perturbation, and the output $y$, which captures the cortical activity recorded from the {EEG} electrodes. It is important to note that the brain source activity $z$ is not directly identifiable through system identification. However, a latent variable $x$ can be estimated. The latent variable, also referred to as the hidden state, is an unobservable variable in a mathematical model, inferred or calculated based on available data to capture underlying patterns or processes that are not directly observable.   

\noindent A linear time-invariant (LTI) mathematical model representing brain dynamics can be expressed as \cite{plub2018state}: 
\begin{equation} \label{eq3:SSmodelBrain}
    \begin{aligned} 
        x(k+1) &= Ax(k) + Bu(k) + w(k), \\
        z(k) &= Hx(k) + \eta(k), \\
        y(k) &= Lz(k) + Du(k) + v(k). 
    \end{aligned}
\end{equation}

\noindent where $x \in \mathbf{R}^n$ denotes the latent variable, $z \in \mathbf{R}^s$ represents the source variables corresponding to cortical activity of $s$ regions, and $y \in \mathbf{R}^{62}$ corresponds to the output representing the electrical activity recorded by the 62 electrodes placed at the scalp during {EEG} measurements, $L \in \mathbf{R}^{{62} \times s}$, and $H \in \mathbf{R}^{s \times n}$. The time interval between two consecutive samples, indexed by $k$ and $k+1$, is $0.488 ms$ in real time. The matrices $A, B, D$, of appropriate size, represent the system matrices, while $w, \eta, v$ are noise terms associated with the latent state, source signal, and output measurements, respectively. The matrix $L$ is the lead field matrix that maps the neural sources to the EEG sensors on the scalp (see Section~\ref{sec:app}). The matrix $H$ represents the mapping from the latent state $x$ to the source activity $z$. 
To identify the system using subspace methods, we first focus on $x, y$ equations, and deal with the source variable $z$ in the next section. To this aim, a modified state-space model, namely the innovation form which takes the measurements and process noise into account\cite{Verhaegen2007FilteringApproach}, is derived from \eqref{eq3:SSmodelBrain} as follows 

\begin{equation} \label{eq3:SSmodelBrain2}
    \begin{aligned} 
        x(k+1) &= Ax(k) + Bu(k) + Ke(k), \\
        y(k) &= Cx(k) + Du(k) + e(k), 
    \end{aligned}
\end{equation}

\noindent where $C = LH$, and $e$ is a zero-mean white noise sequence, and $K$ is the Kalman gain. Our objective is to construct a state-space LTI model that encapsulates the dynamic processes by identifying the matrices $A, B, C, D, K$ such that the following observer estimates the system output closely. That is the approximated $\widehat{y}(k)$ is close enough to $y(k)$. %The implementation of the steady-state Kalman gain $K$ is based on the estimate $\widehat{x}$ of the latent variable:
\begin{equation} \label{eq2:Kalman}
    \begin{aligned} 
    \widehat{x}(k+1) &= A\widehat{x}(k) + Bu(k) + K(y(k) - C\widehat{x}(k) - Du(k)),\\
    \widehat{y}(k) &= C\widehat{x}(k) + Du(k).
    \end{aligned}
\end{equation}

In what follows, we discuss employing two subspace identification algorithms, {N4SID} and {PO-MOESP}, to verify whether an LTI model can closely approximate the EEG measurements. Both methods share the common goal of identifying the state-space model of \eqref{eq3:SSmodelBrain2}, using the input and output data. Here, we first state the underlying assumptions behind the sub-space modeling, and then briefly review the three fundamental stages of both of the aforementioned algorithms. We refer the reader to \cite{Verhaegen2007FilteringApproach} for the detailed theoretical framework and working principles of these algorithms. 
\begin{assumption}\label{ass1}
There exists a minimal, that is observable and controllable, linear time-invariant system that generates the output data.   
\end{assumption}

\begin{assumption}\label{ass3}
The system matrix of the innovation form model in \eqref{eq3:SSmodelBrain2}, $A-KC$, is asymptotically stable.
\end{assumption}

\begin{assumption}\label{ass2}
The system's input is persistently exciting and uncorrelated to the white noise.   
\end{assumption}
\noindent {\bf{PO-MOESP and N4SID algorithms: Overview}}
\begin{enumerate}
\item Data collection and embedding to construct the structured projection, i.e., Hankel and Toeplitz, matrices derived from the provided input and output data,
\item Subspace decomposition by applying matrix decomposition methods such as RQ factorization and singular Value Decomposition (SVD) on the data matrices. In this stage, the model order is determined, and matrices $A$ and $C$ are constructed. We note that the difference of the two subspace methods lies in this step, where SVD of different matrices is used to construct the observability matrix useful in calculating $A$ and $C$. Kalman gain $K$ is also estimated at this stage. 
\item Estimation of matrices $B$ and $D$ by solving a least squares regression problem.
\end{enumerate}
%--------------------------------------------------------------------------------------------------
\subsection{Mapping from latent space to source space}\label{sec:app}
\begin{figure}[h!]
\centering
\includegraphics[width = 0.7\linewidth]{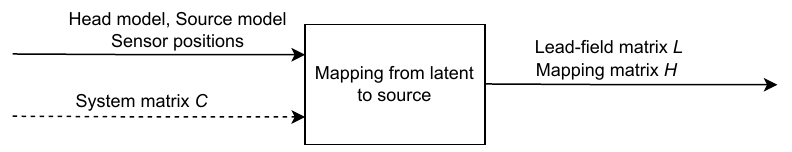}
\caption[Block diagram of step 3]{Block diagram of step 3, illustrating the input and output of the third stage.} 
\label{fig3:step3}
\end{figure}

After obtaining the brain dynamics in the latent space, we can estimate the neural source activities by mapping the latent state variable to source variables. The information of $L$ and $H$, as represented in the model \eqref{eq3:SSmodelBrain2}, is ambiguously mixed in $C = LH$, which is obtained from the subspace identification modeling. In what follows, we provide a concise overview of the manner of construction of matrices $L$ and $H$, and refer the reader to \cite{lim2017sparse,plub2018state} for detailed information. A high-level representation of the construction of $L$ and $H$ is summarized in the block diagram of Figure~\ref{fig3:step3}.

\textbf{Deriving the Lead-field Matrix \(L\):} The lead-field matrix can be estimated from prior knowledge of the head model, source model, and sensor positions of the electrodes on the scalp \cite{Nolte2005AnalyticConductors}\cite{Oostenveld2011FieldTrip:Data}. For the head model we adopted the standard 3-layer Boundary Element Method volume construction model \cite{oostenveld} available in FieldTrip. As the corresponding source model, we used a template-distributed source model representing the grey matter of the highly folded, canonical cortical sheet with 8196 vertices. The individual electrode locations for each participant were warped to the template models before calculating the leadfield. Each source was matched onto the standard automated anatomical labelling (AAL) atlas \cite{Tzourio-Mazoyer2002AutomatedBrain}, consisting of 116 defined brain areas. When mapping the source locations onto the atlas, not all brain areas contained sources, while others contained multiple sources. The brain areas with no sources, 27 in total, were omitted, while for the others, the lead-field matrix values and source grid positions were averaged to create a single representation for each brain area. Following this process, the lead-field matrix $L$ is reduced to dimensions of 62 $\times$ 89, where 62 represents the number of electrodes and 89 denotes the unique brain source locations. No weighting and normalization were applied to the leadfield.\\[1mm]
%As the corresponding source model, we used a template-distributed source model representing the grey matter of the highly folded, canonical cortical sheet with 8196 vertices. The individual electrode locations for each participant were warped to the template models before calculating the leadfield. Each source was matched onto the standard automated anatomical labelling (AAL) atlas  ($"ROI_MNI_V4"$) \cite{Tzourio-Mazoyer2002AutomatedBrain}, consisting of 116 defined brain areas. When mapping source locations onto the atlas, multiple source points correspond to one of the single anatomical regions, specified in the catalogue by Tzourio-Mazoyer et al. \cite{Tzourio-Mazoyer2002AutomatedBrain}. In such instances, the lead-field matrix values and source grid positions are averaged to create a unified representation. Following this process, the lead-field matrix $L$ is reduced to dimensions of 89 $\times$ 62, where 62 represents the number of electrodes and 89 denotes the unique brain source locations. No weighting and normalization were applied to the leadfield.\\[1mm]
\textbf{Deriving the Transformation Matrix \(H\):}
The relationship between \(C\), \(L\), and \(H\) is expressed as:

\begin{equation}
    C = LH
\end{equation}

Since both \(C\) and \(L\) are known, the transformation matrix \(H\) can be computed by solving a linear least squares problem:

\begin{equation}
    \min_H \frac{1}{2}||C - LH||_F^2
\end{equation}

However, due to the under-determined nature of the forward problem, where multiple solutions can fit the data, constraints are imposed on \(H\) to promote group sparsity, reflecting the assumption that only a subset of sources is active at any given time. This can be achieved by formulating the problem as a Group Lasso, optimization with an additional penalty term:

\begin{equation}
\min_H \frac{1}{2}||C - LH||_F^2 + \lambda \sum_{i=1}^{m} ||H_i^T||_2
\end{equation}

Here, \(m\) is the number of sources, and \(\lambda\) is a regularization parameter that controls the degree of sparsity. By adjusting \(\lambda\), we can encourage many rows of \(H\) to be zero, thereby localizing the active brain sources. The optimal sparsity pattern for \(H\) is determined using the Bayesian Information Criterion. The matrix \(H\) selected by minimizing the BIC score to 5 represents the most relevant and active source locations. The matrix $H$ has the dimensions of $89 \times 15$, corresponding to the 89 candidate source locations and 15 latent states. Once $H$ is determined, the source activity $z$ can be computed.

{\bf Quantitative comparison of source activity:} The quantified source activities for the pre- and postcentral areas of the left and right-hemisphere are compared using a paired t-test to explore differences in contra- and ipsilateral brain activation.
%%----------------------------------------------------------------------------------------------------
\section{Results}\label{sec:re}
This section presents the main steps taken to build the linear models, their verification, and validation. We first verify, validate, and compare the performance of the two LTI models obtained from PO-MOESP and N4SID algorithms. For a detailed account of implementation of the algorithms, as well as their corresponding MATLAB codes, we refer to the manual \cite{Verhaegen2007FilteringSoftware}. Second, we establish relation between evolution of latent variables and neural sources and validate the finding of our model by comparing with existing results in the literature.

In the subspace decomposition phase for both methods, the critical task is to determine an appropriate model order $n$. To determine the appropriate model order, an examination of the singular values derived from the subspace decomposition step can be conducted. This involves identifying those values of $n$ that exhibit a significant gap between the singular values $n^{th}$ and $(n+1)^{th}$. In scenarios where multiple such values emerge, the selection process will hinge on a trade-off between achieving good performance, managing computational effort, and avoiding over-fitting. The {PO-MOESP} method, which is more computationally efficient, is employed as the baseline reference. For both {N4SID} and {PO-MOESP}, the same model order is chosen for consistency in comparing the performance. 

When handling extensive datasets, which may encompass data from experiments conducted at different times involving, for example, distinct participants interacting with the same device, it might become imperative to partition the data into separate batches. Subdividing the dataset into smaller batches can be advantageous in training the model on more data, preventing excessive memory usage, and enhancing computational efficiency \cite{Verhaegen2007FilteringSoftware}. Using data batches effectively leverages all the available information from the system. 

We use data from six participants and keep the seventh participant's data for comparison with the output of the obtained model. By computing the {SVD} of one of the sub-matrices of the R matrix, obtained from the RQ factorization of the structured input-output data matrix, the corresponding singular values can be derived. SVD calculations are performed in series. The compressed output from SVD of batch $i$, which encodes the measured data and system dimensions, informs the computation of the SVD for batch $i+1$.

\begin{figure}[h!]
    \centering
    \includegraphics[width = \linewidth, trim=0cm 0cm 0cm 0cm, clip]{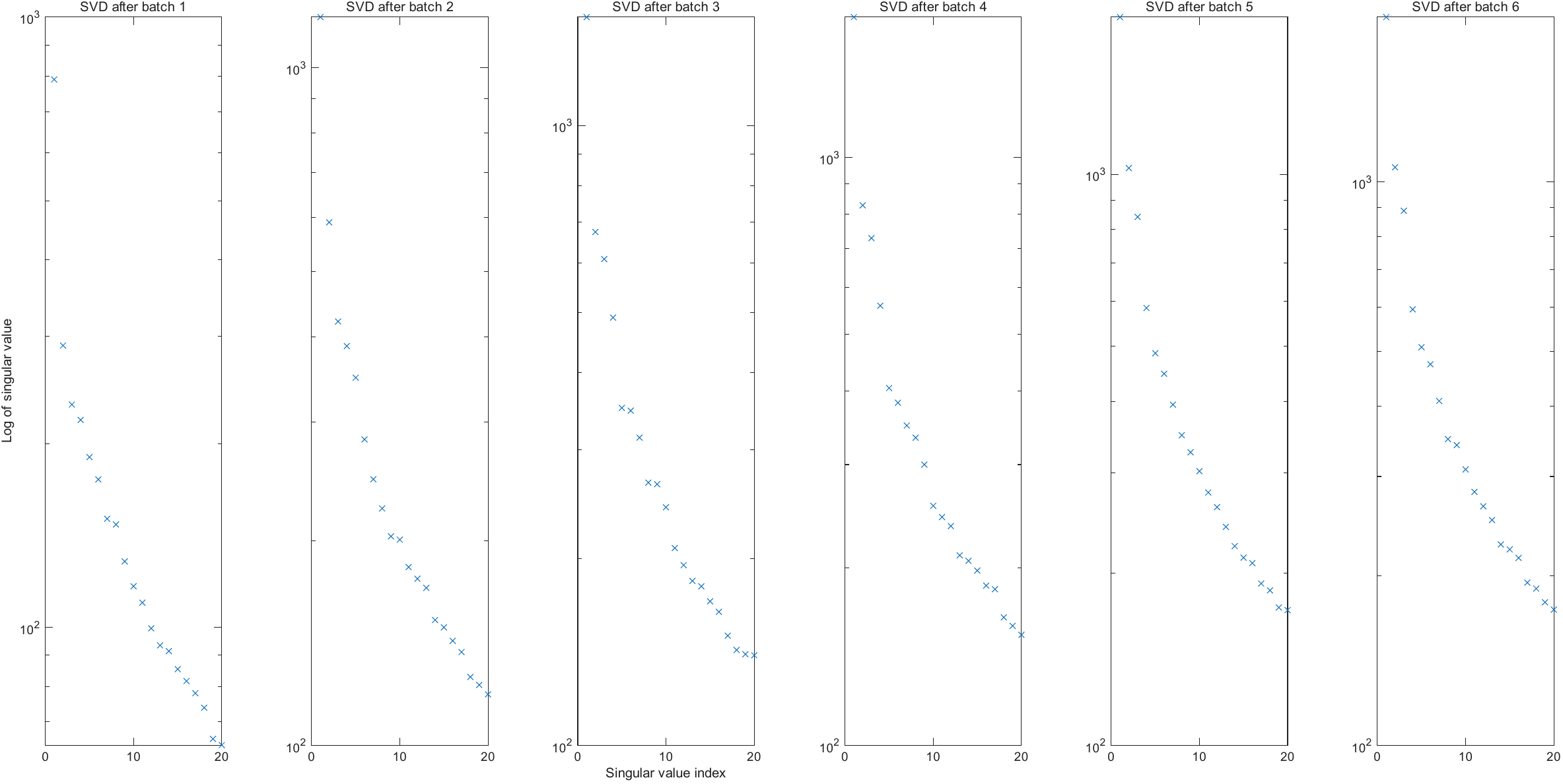}
    \caption {Singular values computed via SVD within the PO-MOESP method, shown after the inclusion of each batch comprising 6 of the 7 participants, for model order determination.}
    \label{fig4:svBatches}
\end{figure}

The singular values obtained after each batch iteration are visualized in Figure \ref{fig4:svBatches}. Examining these singular values across iterations shows a changing pattern in the gaps. Looking at the gaps in the singular values in the plot after the sixth batch, one could choose a model order of $3, 7, 10, 15, 17$ or $18$. 

For each of the above model orders, we obtain the system matrices following the steps $(2)$ and $(3)$ mentioned in the algorithm overview. When multiple significant gaps manifest in the plot of the singular values, the selection of the model order involves a balance between achieving optimal performance, managing computational resources, and preventing overfitting. The variance accounted for (VAF) is a measure of performance that quantifies the proportion of the total variance in the predicted values that can be attributed to the actual values \cite{Verhaegen2007FilteringApproach}. The {VAF} falls within the range of $0\%$ to $100\%$, where higher {VAF} values indicate reduced prediction errors and superior model performance, and is expressed as:
\begin{equation} \label{eq4:VAF}
    \operatorname{VAF}(y(k), \widehat{y}(k))=\max \left(0,\left(1-\frac{\sum_{k=1}^N\|y(k)-\widehat{y}(k)\|_2^2}{\sum_{k=1}^N\|y(k)\|_2^2}\right). \right. %\cdot 100 \%. 
\end{equation}

\noindent In this research, the {VAF} is used to compare the {SSR}, denoted by $y(k)$, and the estimated output using the {PO-MOESP} method, denoted by $\widehat{y}(k)$. These output values represent the electrical activity of a single participant for a specific electrode, measured at a given time instant, $k$. Validated on the data of a single participant, Table~\ref{tab4:orderVAF} represents the mean {VAF} over all the electrodes for different model orders based on the singular values. Notably, at order $n=15$ an increase in {VAF} to $90\%$ occurs compared to lower orders. Elevating the order to 17 or 18 may lead to an increased risk of overfitting and extended computation time without a significant performance improvement. Consequently, for subsequent calculations in both subspace identification methods, the model order $n=15$ is adopted.

\begin{table}[htbp]
\centering
\caption{Selection of model order based on the mean VAF across all electrodes for one participant. Candidate model orders were determined from observed gaps in the singular values of the data matrices constructed from batches of six participants using the PO-MOESP algorithm.}
\begin{tabular}{ll}
\hline
\textbf{Order $n$} & \textbf{VAF {[}$\%${]}} \\ \hline
3                    & 72.26                  \\
7                    & 83.69                   \\
10                   & 87.76                    \\
15                   & 90.45                   \\
17                   & 91.32                   \\
18                   & 91.71              \\ \hline     
\end{tabular}\label{tab4:orderVAF}
\end{table}

The model parameter left to determine is the block size $s$, pertaining to the Hankel matrices, which should be larger than the expected system order and smaller than the number of data points, such that $n < s$. For the ensuing results, $s=20$ meets the criterion \cite{Verhaegen2007FilteringApproach}. 

\subsection{State-space model verification}
Verification of state-space models is crucial to ensure their reliability and validity. Although subspace identification methods can estimate models, they do not automatically guarantee that the models meet essential assumptions \cite{Lacy2003SubspaceOptimization}. In this section, the focus is on testing key assumptions for the reliability of open-loop subspace identification models, including asymptotic stability, observability, and controllability. The stability assumption is verified for models obtained by both methods by examining the eigenvalues of the matrix $A-KC$, which represents the innovation dynamics. The eigenvalues corresponding to both N4SID and PO-MOESP are positioned strictly inside the unit circle, indicating the asymptotic stability of the models.

The final aspect of verification involves the assessment of observability and controllability. Mathematical calculations reveal that the models obtained using both N4SID and PO-MOESP are observable and controllable; that is, the ranks of their observability and controllability matrices are equal to the model order, $n=15$. This alignment with the fundamental requirements for observability and controllability underpins the model's effectiveness in capturing the intricate brain dynamics. It affirms that crucial state variables can be reconstructed from the available output measurements and that the system can be controlled through its inputs.

\subsection{Model performance: Validation and comparison}\label{sc4:validation}
A validation process is conducted to assess the accuracy and reliability of the identified models. The validation criterion measures the model's ability to accurately fit a part of the data that is not employed during the identification process. The evaluation of the model's accuracy is conducted using two key performance metrics, the {VAF}, as defined in equation \eqref{eq4:VAF}, and the {RMSE}. The RMSE assesses the average magnitude of the error, signifying the differences between the
model’s predictions and the actual data points, and is calculated as

\begin{equation}\label{rmse}
RMSE(y(k), \widehat{y}(k))= \sqrt{\frac{\sum_{k=1}^{N} (y(k)-\widehat{y}(k))^2}{N}}.
\end{equation}

To validate the models effectively, the `leave-one-out' cross-validation approach is applied. This involves training the model using data from the first six participants and validating its performance using the data from the seventh participant. This process is iteratively repeated for each of the seven participants, with each participant's dataset serving as the validation set once. The outcomes of the validation process are presented in Table~\ref{tab4:modelPerformance}, which includes the mean and {std} of the performance metrics, expressing the deviations in the performance between the seven participants in the leave-one-out principle. The standard deviation is taken into account to test the robustness of the model. For the model trained with one participant excluded from the training data, participant specific results, i.e., the mean VAF across all electrodes for each participant, using both PO-MOESP and N4SID methods are provided in Table~1 of Supplementary Materials.
\begin{table}[h!]
\centering
\caption{Model performance metrics for all electrodes and participants. In a leave-one-out cross-validation scheme, the mean VAF across all electrodes was calculated for each held-out participant. The reported values correspond to the average of the seven obtained mean VAFs using PO-MOESP and N4SID methods.}
\begin{tabular}{lcccc}
\hline
 & \multicolumn{2}{c}{VAF [\%]} & \multicolumn{2}{c}{RMSE [$\mu$V]} \\
Method & Mean & Std & Mean & Std \\
\hline
N4SID   & 81.55 & 5.20 & 0.26 & 0.05 \\
PO-MOESP & 86.71 & 2.89 & 0.21 & 0.03 \\
\hline
\end{tabular}\label{tab4:modelPerformance}
\end{table}

As presented in Table~\ref{tab4:modelPerformance}, the model validation measures indicate a higher VAF and lower RMSE for the {PO-MOESP} method. The lower standard deviation of the VAF indicates reduced variability in the VAF values obtained across leave-one-out iterations. This suggests that the PO-MOESP method exhibits greater robustness. 

We further validate the estimated output of one of the identified models by comparing the measured signal at electrode CP3 with its corresponding estimated output. Table~\ref{tab:cp3Performance} summarizes the model performance at electrode CP3 for all participants. Participant specific results, i.e., the mean VAF for elctrode CP3 for each participant, using both PO-MOESP and N4SID methods are provided in Table~2 of Supplementary Materials. Figure~\ref{fig4:validation} illustrates the output signals by showing the computed SSR for a representative participant at electrode CP3, alongside the SSR obtained from the reconstructed (estimated) output. These reconstructions are generated using system matrices identified via the N4SID and PO-MOESP methods.
\begin{table}[h!]
\centering
\caption{Model performance metrics for CP3 electrode for all participants. In a leave-one-out cross-validation scheme, the mean VAF of CP3 electrode was calculated for each held-out participant. The reported values correspond to the average of the seven obtained mean VAFs using PO-MOESP and N4SID methods.}
\begin{tabular}{lcccc}
\hline
 & \multicolumn{2}{c}{VAF [\%]} & \multicolumn{2}{c}{RMSE [$\mu$V]} \\
Method & Mean & Std & Mean & Std \\
\hline
N4SID   & 90.18 & 2.61 & 0.24 & 0.06 \\
PO-MOESP & 94.71 & 1.46 & 0.23 & 0.06 \\
\hline
\end{tabular}\label{tab:cp3Performance}
\end{table}

\begin{figure}[h!]
    \centering
    \includegraphics[width = 0.8\linewidth]{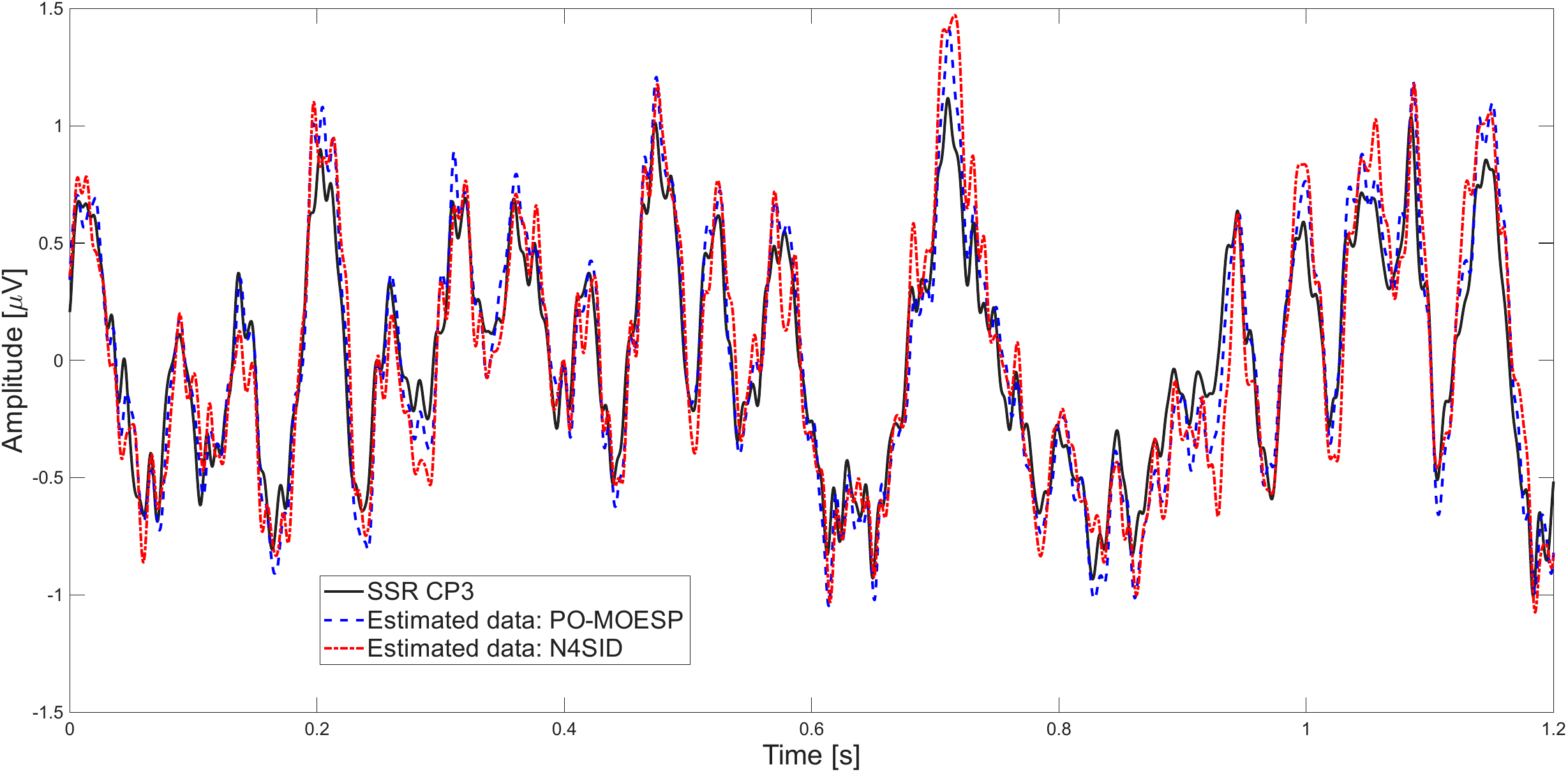}
    \caption[Simulation results of the {SSR} signal]{Comparison of the determined {SSR} at electrode CP3 for a single participant (shown in black), with the reconstructed output estimated by {N4SID} (highlighted in red) and the output reconstructed using the {PO-MOESP} method (shown in blue).}
    \label{fig4:validation}
\end{figure}

Figure~\ref{fig4:validation} shows that the predicted output from the {PO-MOESP} model closely tracks the actual output, displaying a high degree of similarity and fewer offsets compared to the {N4SID} model. This observation is well aligned with the quantitative metrics as reported in Table \ref{tab:cp3Performance}. The obtained model allows us to describe the brain's cortical network linearly in a latent space. In the next section, we explore the possibility of an LTI model whose state variables are the neural source activities.\\[1mm]
\noindent {\bf Remark:} As demonstrated, by employing two classical systems and control theory–oriented subspace identification methods, we have shown that a linear state-space model is capable of capturing the latent dynamics of EEG signals during the passive motor task. We note that, with the continued development of system identification techniques, a wide range of alternative methods is available. However, the objective of this work is not to compare identification techniques in terms of performance, but rather to investigate the nature of EEG-driven dynamics in the presence of exogenous inputs for the target task, and to support this objective using two methods.
%%%%%%%%%%%%%%%%%%%%%%%%%%%%%%%%%%%%%%%%%%%%%%%%%%%%%%%%%%%%%%%%%%%%%%%%%%%%%%%%%%%%%%%%%%%%%%%%%%%%%%%%%%%%%%%%%%%%%%%%%%%%%%%%%%%%%%%%
\subsection{Validation in the source space}
We now discuss translating the latent states $x$, reconstructed in the previous phase, to the neural source states represented by $z$, enabling the extraction of insights into the interconnections between brain sources and regions. We construct two matrices $L$ and $H$, as explained in Section \ref{sec:app}, to describe the relationship between the electrical activity generated by the neural sources of the brain and the EEG measurements, and to map the latent variables to the brain's sources.

The results for the source activity of a single participant are shown in Figure~\ref{fig4:sourceReconstruction}. Results for all participants are presented in the Supplementary Materials, Figure~1. For visualization purposes, only the ten sources with the greatest activity across the brain are shown, all of which happen to be located on the left side of the brain. Both the color and size indicate the relative strengths of the sources. The greatest source activity is observed for the left cortical sensorimotor structures expected to be active while the right wrist is passively moved. A paired t-test confirmed that the source activity in the left pre- and postcentral areas is significantly larger than on the right side (precentral - Left: 9258.6 $\pm$ 3484.6 a.u.; Right: 1235.5 $\pm$ 648.4 a.u.; $t(6) = 7.19, p<0.001$; and postcentral - Left: 8424.1 $\pm$ 3153.0 a.u.; Right: 1360.8 $\pm$ 444.8 a.u.; $t(6) = 6.51, p<0.001)$. The observed source activities are in line with the electrode-level observations in \cite{Vlaar2017QuantificationStroke}. They reported the highest electrode-level SNR over the contralateral somatosensory central and parietal areas, whereas the ipsilateral somatosensory areas presented a low SNR. A high SNR is indicative of an SSR with little variance over epochs, but does not directly provide information on the absolute SSR amplitude. This finding is translated to the source space, where we find the greatest source activity contralateral to the wrist that is passively being moved. 
In the employed experimental setup, the wrist of the human participants is passively moved while the hand is attached to a manipulator's handle. This provokes a continuous stream of proprioceptive and tactile sensory information to the cerebral cortex. Mechanoreceptors in the skin convey their information over dorsal column-medial lemniscal pathway to the contralateral primary somatosensory and motor cortex after being relayed in the thalamus. The primary somatosensory cortex distributes information further over to the secondary somatosensory and posterior parietal cortices \cite{Chang2009TheStudy,Vakorin2010ComplexityResponse,szameitat2012cortical}. The source localisation results of this study are consistent with these cortical areas, and are similar to cortical activation patterns previously reported for mechanically evoked SSR \cite{Vlaar2017QuantificationStroke,severens,langdon}.
\begin{figure}[h!]
    \centering
    \includegraphics[width = 0.55\linewidth]{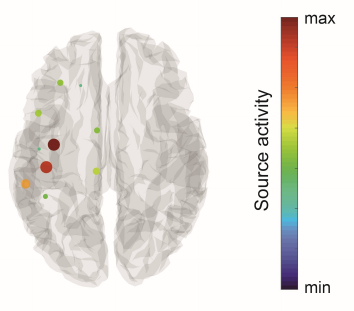}
\caption{Estimated source activity for a single participant. The ten sources with the greatest source activity for this participant are shown. Both the color and size indicate the relative strengths of the sources. The largest source activity is found over the pre- and postcentral area of the left hemisphere – contralateral to the side of the imposed joint perturbation.}    
    \label{fig4:sourceReconstruction}
\end{figure} 
%%%%%%%%%%%%%%%%%%%%%%%%%%%%%%%%%%%%%%%%%%%%%%%%%%%%%%%%%%%%%%%%%%%%%%%%%%%%%%%%%%%%%%%%%%%%%%%%%%%%%%%%%%%%%%%%%%%%%%%%%%%%%%%%%%%%%%%%%%%%%%
\section{Discussion}\label{sec:dis}
In the previous sections, we have demonstrated that brain dynamics, as observed in EEG data from a passive motor task, can be described by a linear state-space model. However, this model is not unique \cite{vlaar2016quantifying}, as the resulting representation depends on several factors, including the modeling technique. In this section, we elaborate on this finding and discuss its potential applications and limitations. 

\subsection*{Linear macroscopic dynamics of the brain experiencing exogenous inputs}
In general, time-domain modeling of the brain has been explored using two main approaches: parameter identification for models with predefined structures and data-driven models. The former includes several known computational models ranging from mean-field and neural mass models to detailed spiking neuronal networks \cite{gerstner2014neuronal}. These classes of models, mainly nonlinear differential equations, can incorporate known biophysical phenomena in addition to measured data, resulting in a high level of biophysical detail. However, this level of detail increases computational and analytical complexity, which grows with the size and complexity of the involved brain regions. In contrast, data-driven modeling, does not include such fine details and mainly focuses on capturing the high-level dynamics of information exchange. However, data-driven approaches can also face substantial computational and analytical complexity, particularly when nonlinear differential equations are employed \cite{friston2003dynamic}.   

While nonlinear dynamics allow incorporation of biological detail, they often increase computational and analytical complexity. Recent work has examined the potential of linear models to capture brain dynamics, primarily in the resting state; notably, \cite{nozari2024macroscopic} provided evidence that linear time-invariant models accurately describe macroscopic resting-state dynamics. However, exogenous inputs can reveal additional dynamical features and thus may affect these conclusions. Our results extend the findings in \cite{nozari2024macroscopic} by showing that macroscopic dynamics during a passive motor task, in the presence of exogenous input, are well captured by linear time-invariant state-space models. As discussed in the paper, we have validated these models in both latent and source spaces.

\subsection*{Limitations and considerations}
Given advances in measurement and stimulation, revealing brain dynamics and designing simple, efficient controllers have become increasingly important. Model complexity depends on modeling and control objectives and on task complexity. Determining whether linear models can capture cortical dynamics requires data from complex tasks. We studied a passive motor task and used whole‑brain EEG; results may not generalize to more complex tasks or to recordings with fewer sensors. Thus, while linear state‑space models offer a useful first approximation of macroscopic brain dynamics, their validity should be tested for complex tasks and number of sensed cortical regions.

\subsection*{Potentials}
\begin{figure}[h!]
\centering
    \includegraphics[width = 0.7\linewidth]{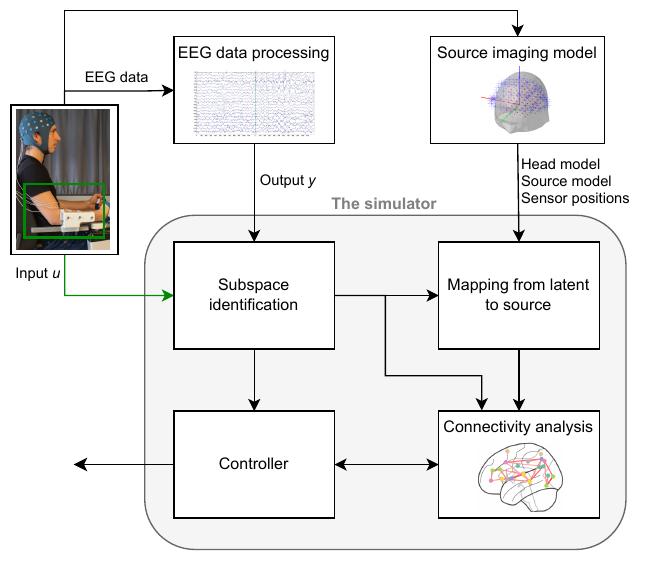}
    \caption{Schematic of Simulator and control design}
    \label{fig:app}
\end{figure}

Linear state-space models are useful in enabling the design of tractable controllers and providing the possibility of conducting connectivity analysis for brain networks. They also permit tracking changes in brain sub-networks, for example, by estimating time-varying connectivity or state dynamics by calculating system responses to various inputs. Our results provide a basis for developing simulation and control-design environments in which modeling, connectivity analysis \cite{Barnett2015GrangerModels}, and control design are feasi, as shown in Figure ~\ref{fig:app}. %The control design could take place sequentially or simultaneously with identification. The latter deserves further studies. 
The future perspective of our results include providing a new avenue for exploring the fundamental mechanisms underlying the brain's processing of external sensory stimuli. In case of neurological dysfunction, such as following stroke, Parkinson’s disease, and Alzheimer’s disease, disruptions in the processing of sensory stimuli are often present. Information from the state-space model not only allows for the visualization of changes in connectivity over time but also enables the design of controllers to aid in managing revealed deficits when needed. This could, for example, be achieved through the improvement of hybrid brain-computer interfaces that use external sensory stimuli for control.

%---------------------------------------
\subsection{Conclusion}\label{sec:co}
We have employed subspace identification algorithms to present a linear time-invariant model describing the generative dynamics of electroencephalography (EEG) data in an experiment with continuous exogenous input. The EEG data were recorded while participants were seated with their wrists strapped to a haptic manipulator. We have shown that the LTI model obtained using the PO-MOESP algorithm accurately represents the data in a latent space. Furthermore, we verified the obtained linear model in computing the activities of brain regions in the source space. We discussed the usefulness of our approach in connectivity analysis and control design to treat impairments in which robotic assistance or brain-interface devices are applicable. Further avenues include closed-loop identification and stimulation and verification of the approach in more complex situations.

% --- Acknowledgement section ---
\section*{Acknowledgement}
The authors would like to thank Alfred Schouten for providing the video of the experiment, and Michel Verhaegen for helpful discussions. This research was supported by the Netherlands Organization of Scientific Research (NWO) under the gravitation project, Dutch Brain Interface Initiative DBI2.

\bibliographystyle{plain}  % choose the style
\bibliography{references} 

@article{Bell1995AnDeconvolution.,
    title = {{An information-maximization approach to blind separation and blind deconvolution.}},
    year = {1995},
    journal = {Neural computation},
    author = {Bell, Anthony J. and Sejnowski, Terrence J.},
    number = {6},
    volume = {7},
    doi = {10.1162/neco.1995.7.6.1129},
    issn = {08997667}
}

@article{Nolte2005AnalyticConductors,
    title = {{Analytic Expansion of the EEG Lead Field for Realistic Volume Conductors}},
    year = {2005},
    journal = {Physics in Medicine and Biology},
    author = {Nolte, Guido and Dassios, George},
    number = {16},
    volume = {50},
    doi = {10.1088/0031-9155/50/16/010},
    issn = {00319155}
}

@article{Tzourio-Mazoyer2002AutomatedBrain,
    title = {{Automated Anatomical Labeling of Activations in SPM Using a Macroscopic Anatomical Parcellation of the MNI MRI Single-Subject Brain}},
    year = {2002},
    journal = {NeuroImage},
    author = {Tzourio-Mazoyer, N. and Landeau, B. and Papathanassiou, D. and Crivello, F. and Etard, O. and Delcroix, N. and Mazoyer, B. and Joliot, M.},
    number = {1},
    volume = {15},
    doi = {10.1006/nimg.2001.0978},
    issn = {10538119}
}

@article{Vakorin2010ComplexityResponse,
    title = {{Complexity Analysis of Source Activity Underlying the Neuromagnetic Somatosensory Steady-State Response}},
    year = {2010},
    journal = {NeuroImage},
    author = {Vakorin, Vasily A. and Ross, Bernhard and Krakovska, Olga and Bardouille, Timothy and Cheyne, Douglas and McIntosh, Anthony R.},
    number = {1},
    volume = {51},
    doi = {10.1016/j.neuroimage.2010.01.100},
    issn = {10538119}
}

@article{Breakspear2017DynamicActivity,
    title = {{Dynamic Models of Large-Scale Brain Activity}},
    year = {2017},
    journal = {Nature Neuroscience},
    author = {Breakspear, Michael},
    number = {3},
    month = {2},
    pages = {340--352},
    volume = {20},
    publisher = {Nature Publishing Group},
    doi = {10.1038/NN.4497},
    issn = {15461726},
    pmid = {28230845}
}

@article{Delorme2004EEGLAB:Analysis,
    title = {{EEGLAB: An Open Source Toolbox for Analysis of Single-Trial EEG Dynamics including Independent Component Analysis}},
    year = {2004},
    journal = {Journal of Neuroscience Methods},
    author = {Delorme, Arnaud and Makeig, Scott},
    number = {1},
    month = {3},
    pages = {9--21},
    volume = {134},
    publisher = {Elsevier},
    doi = {10.1016/J.JNEUMETH.2003.10.009},
    issn = {01650270},
    pmid = {15102499}
}

@article{Cheung2010EstimationModels,
    title = {{Estimation of Cortical Connectivity from EEG Using State-Space Models}},
    year = {2010},
    journal = {IEEE Transactions on Biomedical Engineering},
    author = {Cheung, Bing Leung Patrick and Riedner, Brady Alexander and Tononi, Giulio and Van Veen, Barry D.},
    number = {9},
    month = {9},
    pages = {2122--2134},
    volume = {57},
    doi = {10.1109/TBME.2010.2050319},
    issn = {00189294},
    pmid = {20501341}
}

@article{Oostenveld2011FieldTrip:Data,
    title = {{FieldTrip: Open Source Software for Advanced Analysis of MEG, EEG, and Invasive Electrophysiological Data}},
    year = {2011},
    journal = {Computational Intelligence and Neuroscience},
    author = {Oostenveld, Robert and Fries, Pascal and Maris, Eric and Schoffelen, Jan Mathijs},
    volume = {2011},
    publisher = {Hindawi Limited},
    doi = {10.1155/2011/156869},
    issn = {16875273},
    pmid = {21253357}
}

@book{Verhaegen2007FilteringApproach,
  title={Filtering and system identification: A least squares approach},
  author={Verhaegen, Michel and Verdult, Vincent},
  year={2007},
  publisher={Cambridge university press}
}

@article{Verhaegen2007FilteringSoftware,
    title = {{Filtering and System Identification: An Introduction to using Matlab Software}},
    year = {2007},
    journal = {Delft University of Technology},
    author = {Verhaegen, Michel and Verdult, Vincent and Bergboer, Niek},
    number = {163},
    volume = {68},
    url = {https://www.dcsc.tudelft.nl/~jwvanwingerden/lti/doc/companion.pdf}
}

@article{Barnett2015GrangerModels,
    title = {{Granger Causality for State-Space Models}},
    year = {2015},
    author = {Barnett, Lionel and Seth, Anil K},
    arxivId = {1501.06502v2}
}

@article{Manomaisaowapak2022GrangerApproach,
    title = {{Granger Causality Inference in EEG Source Connectivity Analysis: A State-Space Approach}},
    year = {2022},
    journal = {IEEE Transactions on Neural Networks and Learning Systems},
    author = {Manomaisaowapak, Parinthorn and Nartkulpat, Anawat and Songsiri, Jitkomut},
    number = {7},
    month = {7},
    pages = {3146--3156},
    volume = {33},
    publisher = {Institute of Electrical and Electronics Engineers Inc.},
    doi = {10.1109/TNNLS.2021.3096642},
    issn = {21622388},
    pmid = {34310324}
}

@article{Do2021HumanNavigation,
    title = {{Human Brain Dynamics in Active Spatial Navigation}},
    year = {2021},
    journal = {Scientific Reports 2021 11:1},
    author = {Do, Tien Thong Nguyen and Lin, Chin Teng and Gramann, Klaus},
    number = {1},
    month = {6},
    pages = {1--12},
    volume = {11},
    publisher = {Nature Publishing Group},
    url = {https://www.nature.com/articles/s41598-021-92246-4},
    isbn = {0123456789},
    doi = {10.1038/s41598-021-92246-4},
    issn = {2045-2322},
    pmid = {34158525}
}

@article{Becker2018Large-ScaleIdentification,
    title = {{Large-Scale Dynamic Modeling of Task-fMRI Signals via Subspace System Identification}},
    year = {2018},
    journal = {Journal of neural engineering},
    author = {Becker, Cassiano O. and Bassett, Danielle S. and Preciado, Victor M.},
    number = {6},
    month = {9},
    volume = {15},
    publisher = {J Neural Eng},
    url = {https://pubmed.ncbi.nlm.nih.gov/30088476/},
    doi = {10.1088/1741-2552/AAD8C7},
    issn = {1741-2552},
    pmid = {30088476}
}

@article{Ubeyli2010LeastSignals,
    title = {{Least Squares Support Vector Machine Employing Model-Based Methods Coefficients for Analysis of EEG Signals}},
    year = {2010},
    journal = {Expert Systems with Applications},
    author = {{\"{U}}beyli, Elif Derya},
    number = {1},
    month = {1},
    pages = {233 - 239},
    volume = {37},
    doi = {10.1016/j.eswa.2009.05.012},
    issn = {09574174}
}

@article{Anderson1998MultivariateTasks,
    title = {{Multivariate Autoregressive Models for Classification of Spontaneous Electroencephalographic Signals during Mental Tasks}},
    year = {1998},
    journal = {IEEE Transactions on Biomedical Engineering},
    author = {Anderson, Charles W. and Stolz, Erik A. and Shamsunder, Sanyogita},
    number = {3},
    month = {3},
    pages = {277 - 286},
    volume = {45},
    publisher = {IEEE},
    doi = {10.1109/10.661153},
    issn = {00189294},
    pmid = {9509744}
}

@article{VanOverschee1994N4SID:Systems,
    title = {{N4SID: Subspace Algorithms for the Identification of Combined Deterministic-Stochastic Systems}},
    year = {1994},
    journal = {Automatica},
    author = {Van Overschee, Peter and De Moor, Bart},
    number = {1},
    month = {1},
    pages = {75--93},
    volume = {30},
    publisher = {Pergamon},
    doi = {10.1016/0005-1098(94)90230-5},
    issn = {0005-1098}
}

@article{Vlaar2017QuantificationStroke,
    title = {{Quantification of Task-Dependent Cortical Activation Evoked by Robotic Continuous Wrist Joint Manipulation in Chronic Hemiparetic Stroke}},
    year = {2017},
    journal = {Journal of NeuroEngineering and Rehabilitation},
    author = {Vlaar, Martijn P. and Solis-Escalante, Teodoro and Dewald, Julius P.A. and Van Wegen, Erwin E.H. and Schouten, Alfred C. and Kwakkel, Gert and Van Der Helm, Frans C.T.},
    number = {1},
    month = {4},
    pages = {1--15},
    volume = {14},
    publisher = {BioMed Central Ltd.},
    url = {https://jneuroengrehab.biomedcentral.com/articles/10.1186/s12984-017-0240-3},
    doi = {10.1186/S12984-017-0240-3/FIGURES/3},
    issn = {17430003},
    pmid = {28412953}
}

@article{Chisci2010Real-TimeMachines,
    title = {{Real-Time Epileptic Seizure Prediction Using AR Models and Support Vector Machines}},
    year = {2010},
    journal = {IEEE Transactions on Biomedical Engineering},
    author = {Chisci, Luigi and Mavino, Antonio and Perferi, Guido and Sciandrone, Marco and Anile, Carmelo and Colicchio, Gabriella and Fuggetta, Filomena},
    number = {5},
    month = {5},
    pages = {1124 - 1132},
    volume = {57},
    doi = {10.1109/TBME.2009.2038990},
    issn = {00189294},
    pmid = {20172805}
}

@article{Parra2005RecipesEEG,
  title={Recipes for the linear analysis of EEG},
  author={Parra, Lucas C and Spence, Clay D and Gerson, Adam D and Sajda, Paul},
  journal={Neuroimage},
  volume={28},
  number={2},
  pages={326--341},
  year={2005},
  publisher={Elsevier}
}

@article{Lacy2003SubspaceOptimization,
    title = {{Subspace identification with guaranteed stability using constrained optimization}},
    year = {2003},
    journal = {IEEE Transactions on Automatic Control},
    author = {Lacy, Seth L. and Bernstein, Dennis S.},
    number = {7},
    volume = {48},
    doi = {10.1109/TAC.2003.814273},
    issn = {00189286}
}

@article{Chang2009TheStudy,
    title = {{The Comparison of Cortical Activation Patterns by Active Exercise, Proprioceptive Input, and Touch Stimulation in the Human Brain: A Functional MRI Study}},
    year = {2009},
    journal = {NeuroRehabilitation},
    author = {Chang, Min Chul and Ahn, Sang Ho and Cho, Yun Woo and Son, Su Min and Kwon, Yong Hyun and Lee, Mi Young and Byun, Woo Mok and Jang, Sung Ho},
    number = {2},
    volume = {25},
    doi = {10.3233/NRE-2009-0502},
    issn = {10538135}
}

@article{vanKordelaar2021TheStroke,
    title = {{The Cortical Response Evoked by Robotic Wrist Perturbations Reflects Level of Proprioceptive Impairment After Stroke}},
    year = {2021},
    journal = {Frontiers in Human Neuroscience},
    author = {van Kordelaar, Joost and van de Ruit, Mark and Solis-Escalante, Teodoro and Aerden, Leo A.M. and Meskers, Carel G.M. and van Wegen, Erwin E.H. and Schouten, Alfred C. and Kwakkel, Gert and van der Helm, Frans C.T.},
    month = {11},
    pages = {695366},
    volume = {15},
    publisher = {Frontiers Media S.A.},
    doi = {10.3389/FNHUM.2021.695366/BIBTEX},
    issn = {16625161}
}

@article{Michel2012TowardsTool,
    title = {{Towards the Utilization of EEG as a Brain Imaging Tool}},
    year = {2012},
    journal = {NeuroImage},
    author = {Michel, Christoph M. and Murray, Micah M.},
    number = {2},
    month = {6},
    pages = {371--385},
    volume = {61},
    publisher = {Academic Press},
    doi = {10.1016/J.NEUROIMAGE.2011.12.039},
    issn = {1053-8119},
    pmid = {22227136}
}

@article{szameitat2012cortical,
  title={Cortical activation during executed, imagined, observed, and passive wrist movements in healthy volunteers and stroke patients},
  author={Szameitat, A.J. and Shen, S. and Conforto, A. and Sterr, A.},
  journal={Neuroimage},
  volume={62},
  number={1},
  pages={266--280},
  year={2012},
  publisher={Elsevier}
}

@article{lim2017sparse,
  title={Sparse EEG/MEG source estimation via a group lasso},
  author={Lim, M. and Ales, J.M. and Cottereau, B.R. and Hastie, T. and Norcia, A.M.},
  journal={PloS one},
  volume={12},
  number={6},
  pages={e0176835},
  year={2017},
  publisher={Public Library of Science San Francisco, CA USA}
}

@inproceedings{plub2018state,
  title={State-space model estimation of EEG time series for classifying active brain sources},
  author={Plub-in, N. and Songsiri, J.},
  booktitle={2018 11th Biomedical Engineering International Conference},
  pages={1--5},
  year={2018},
  organization={IEEE}
}

@article{vakitbilir2025multivariate,
  title={Multivariate linear time-series modeling and prediction of cerebral physiologic signals: review of statistical models and implications for human signal analytics},
  author={Vakitbilir, N. and Sainbhi, A. S. and Islam, A. and Gomez, A. and Stein, K.Y. and Froese, L. and Bergmann, T. and McClarty, D. and Raj, R. and Zeiler, F. A.},
  journal={Frontiers in Network Physiology},
  volume={5},
  pages={1551043},
  year={2025},
  publisher={Frontiers Media SA}
}

@article{nozari2024macroscopic,
  title={Macroscopic resting-state brain dynamics are best described by linear models},
  author={Nozari, E. and Bertolero, M.A. and Stiso, J. and Caciagli, L. and Cornblath, E.J. and He, X. and Mahadevan, A.S. and Pappas, G.J. and Bassett, D.S.},
  journal={Nature biomedical engineering},
  volume={8},
  number={1},
  pages={68--84},
  year={2024},
  publisher={Nature Publishing Group UK London}
}

@article{friston2003dynamic,
  title={Dynamic causal modelling},
  author={Friston, K.J. and Harrison, L. and Penny, W.},
  journal={Neuroimage},
  volume={19},
  number={4},
  pages={1273--1302},
  year={2003},
  publisher={Elsevier}
}

@book{gerstner2014neuronal,
  title={Neuronal dynamics: From single neurons to networks and models of cognition},
  author={Gerstner, W. and Kistler, W.M. and Naud, R. and Paninski, L.},
  year={2014},
  publisher={Cambridge University Press}
}

@article{oostenveld,
  title={Brain symmetry and topographic analysis of lateralized event-related potentials},
  author={Oostenveld, Robert and Stegeman, Dick F and Praamstra, Peter and van Oosterom, Adriaan},
  journal={Clinical neurophysiology},
  volume={114},
  number={7},
  pages={1194--1202},
  year={2003},
  publisher={Elsevier}
}

@book{pintelon2012system,
  title={System identification: a frequency domain approach},
  author={Pintelon, R. and Schoukens, J.},
  year={2012},
  publisher={John Wiley \& Sons}
}

@article{pion,
  title={ICLabel: An automated electroencephalographic independent component classifier, dataset, and website},
  author={Pion-Tonachini, Luca and Kreutz-Delgado, Ken and Makeig, Scott},
  journal={NeuroImage},
  volume={198},
  pages={181--197},
  year={2019},
  publisher={Elsevier}
}

@article{langdon,
  title={Multi-frequency phase locking in human somatosensory cortex},
  author={Langdon, Angela J and Boonstra, Tjeerd W and Breakspear, Michael},
  journal={Progress in biophysics and molecular biology},
  volume={105},
  number={1-2},
  pages={58--66},
  year={2011},
  publisher={Elsevier}
}

@article{severens,
  title={A multi-signature brain--computer interface: use of transient and steady-state responses},
  author={Severens, Marianne and Farquhar, Jason and Duysens, Jacques and Desain, Peter},
  journal={Journal of neural engineering},
  volume={10},
  number={2},
  pages={026005},
  year={2013},
  publisher={IOP Publishing}
}

@article{vlaar2016quantifying,
  title={Quantifying nonlinear contributions to cortical responses evoked by continuous wrist manipulation},
  author={Vlaar, Martijn P and Solis-Escalante, Teodoro and Vardy, Alistair N and Van Der Helm, Frans CT and Schouten, Alfred C},
  journal={IEEE transactions on neural systems and rehabilitation engineering},
  volume={25},
  number={5},
  pages={481--491},
  year={2016},
  publisher={IEEE}
}

@article{dziarnowska2026associative,
  title={Associative emotional memory encoding: insights from network stability analysis of an fMRI-driven bilinear dynamics},
  author={Dziarnowska, Weronika and Orhun, Melis and Zhu, Yannan and Kohn, Nils and Fern{\'a}ndez, Guill{\'e}n and Jafarian, Matin},
  journal={Frontiers in Systems Neuroscience},
  volume={20},
  pages={1794170},
  year={2026},
  publisher={Frontiers Media SA}
}
\end{document}